\def\mass#1{${\mathrm{#1\,M}_\odot}$}
\def\mmass#1{{\mathrm{#1\,M}_\odot}}
\def\chem#1#2{$\mathrm{^{#2}\kern-0.8pt#1}$}
\def\mchem#1#2{\mathrm{^{#2}\kern-0.8pt#1}}
\def\reac#1#2#3#4#5#6{$\mathrm{\,^{#2}\kern-0.8pt{#1}\,({#3}\,,{#4})\,{}^{#6}\kern-0.8pt{#5}\,}$}
\def\betap#1#2#3#4{$\mathrm{\,^{#2}\kern-0.8pt{#1}\,(\beta^+)\,{}^{#4}\kern-0.8pt{#3}\,}$}
\def\betam#1#2#3#4{$\mathrm{\,^{#2}\kern-0.8pt{#1}\,(\beta^-)\,{}^{#4}\kern-0.8pt{#3}\,}$}
\def\reacbp#1#2#3#4#5#6#7#8{$\mathrm{\,^{#2}\kern-0.8pt{#1}\,({#3}\,,{#4})\,{}^{#6}\kern-0.8pt{#5}\,(\beta^+)\,{}^{#8}\kern-0.8pt{#7}\,}$}
\def\reacbm#1#2#3#4#5#6#7#8{$\mathrm{\,^{#2}\kern-0.8pt{#1}\,({#3}\,,{#4})\,{}^{#6}\kern-0.8pt{#5}\,(\beta^-)\,{}^{#8}\kern-0.8pt{#7}\,}$}
\def\simgr{\mathbin{\;\raise1pt\hbox{$>$}\kern-8pt\lower3pt\hbox{$\sim$}\;}}
\def\simlr{\mathbin{\;\raise1pt\hbox{$<$}\kern-8pt\lower3pt\hbox{$\sim$}\;}}
\def\ontops#1#2{^{#1}_{#2}}
\def\ontop#1#2{\begin{array}{c}
               #1 \\ #2
               \end{array}}

\documentclass{aa}
\usepackage{graphics}
\begin{document}

\thesaurus{19.34.1; 19.38.1; 19.41.1; 07.25.1; 19.69.1}

\title{On some properties of very metal-rich stars}
%\subtitle{}         % is optional
 
\author{ N. Mowlavi$^1$\ \and
         G. Meynet$^1$\ \and
         A. Maeder$^1$\ \and
         D. Schaerer$^2$\ \and
         C. Charbonnel$^3$\
       }
                                
% \offprints{ }          % is optional

\institute{$1$ Geneva Observatory, CH-1290 Sauverny, Switzerland\\
           $2$ Space Telescope Science Institute, 3700 San Martin Drive,
Baltimore, MD 21218, USA\\
           $3$ Laboratoire d'Astrophysique de Toulouse, CNRS - UMR 5572, 14 av.
E. Belin, 31400 Toulouse, France}

\date{Received 30 September 1997; accepted 6 April 1998}

\maketitle

\begin{abstract}
  An analysis of some properties of stellar models as a function of
metallicity $Z$ (and helium content $Y$) is presented, with special attention to
those stars with metallicities higher than twice or three times solar.

  It is shown that the stellar properties as a function of $Z$ are
mainly determined by the effects of the opacities at sub-solar
metallicities, and by the effects of the mean molecular weight and stellar
mass loss at higher metallicities.
As a result, very metal-rich stars ($Z\simgr 0.05$) exhibit
properties that deviate from what is expected from the known characteristics at
lower metallicities.
  In particular, they are more luminous and hotter than those
at $Z\simlr 0.05$ due to the effect of the mean molecular weight. They
have main sequence lifetimes much shorter (60\% shorter at $Z=0.1$ than at $Z=0.02$)
than those at solar metallicity due
to their lower initial hydrogen content.
Finally, the high mass loss rates at high metallicities affect significantly
the population synthesis of massive stars in very metal-rich regions.
An analysis of expected Wolf-Rayet and supernovae populations in such conditions
is briefly presented.

\end{abstract}

\section{Introduction}

Observations as well as models of galactic chemical evolution
suggest that the metallicity at the center of our galaxy, where the
density of
matter is high, can reach up to 3 to 5 times the solar metallicity (McWilliam
\&
Rich 1994, Simpson et al. 1995). Other objects such as elliptical
galaxies, or maybe even quasars (Korista et al. 1996), also show evidences of
very metal-rich stellar populations.

  The metallicity affects the evolution of stars mainly through its
impact on the radiative opacities, the equation of state, the nuclear
reaction rates and the stellar mass loss rates.
The impact of the metallicity on the stellar properties has been well
studied for $0.001\le Z \le 0.04$ on base of extensive stellar model
calculations
(see e.g. Schaller et al. 1992, hereafter Paper I, and Maeder \& Conti 1994).
In contrast, they are still poorly known at metallicities much higher than
solar.
Recent calculations have been performed at metallicities five times solar
(Fagotto et al. 1994, Mowlavi et al.  1997). Mowlavi et al. (1997) notice that
those very metal-rich models present specific properties which are distinct
from the ones at $Z\le 0.04$. In particular, they are {\it hotter} and {\it more
luminous} that those at solar metallicity, contrary to expectations.
It is the purpose of this paper to analyze in detail some of
these properties of very metal-rich stars. These could have important
consequences on the observable properties of high metal-rich galaxies.

The study is based on the extensive grids of stellar models computed
by the Geneva group with $0.001 \le Z \le 0.1$. It covers the main phases of
stellar evolution, excluding the latest stages of evolution. The core helium
burning phase of very metal-rich low-mass ($M<\mmass{1.7}$) stars, important for
the study of elliptical
galaxies, is not considered in this paper as it has been the subject of
existing papers in the literature (see, for example, Dorman, Rood \& O'Connell,
1993, Fagotto et al. 1994).

A general discussion on the dependence of stellar properties on chemical
composition is presented in Sect.~\ref{Sect:chemical composition}.
Section \ref{Sect:models} summarizes the model ingredients used in the
computation of the Geneva grids. The stellar properties during the main sequence
(MS) and the core helium burning (CHeB) phase are analyzed in Sects.~\ref{Sect:MS} and
\ref{Sect:CHeB}, respectively.
Wolf-Rayet stars in very metal-rich environments are briefly
discussed in Sect.~\ref{Sect:WR}.  Conclusions are drawn in
Sect.~\ref{Sect:conclusions}.

\section{Chemical composition and stellar properties}
\label{Sect:chemical composition}

  Chemical composition is traditionally described by $X$, $Y$ and $Z$, which are the mass
fractions of hydrogen, helium and heavier elements, respectively. In the course of galactic
evolution, the interstellar matter is progressively enriched in heavy elements,
and $Z$ increases through time. It is evident that helium is also enriched, the
increase of its interstellar abundance with respect to $Z$ being conveniently described by a
$\Delta Y/\Delta Z$ law such that $Y=Y_0+\Delta
Y/\Delta Z\times Z$ (where $Y_0$ is the primordial He abundance). At low metallicities
($Z\simlr 0.02$),
$Y$ remains practically constant [an absolute variation $\delta Z$ of the
metallicity induces an absolute variation of He of $\delta Y = \Delta Y/\Delta Z \times \delta Z$,
so that $\delta Y/Y \simlr 10^{-2}$ for $Z\simlr 10^{-3}$], and the
stellar properties {\it as a function of metallicity} are thus expected to be
determined mainly by $Z$. At high metallicities ($Z\simgr 0.02$), on
the other hand, $Y$ increases (or, alternatively, $X$ decreases) significantly
with $Z$ ($Y$ passes from 0.30 to 0.48 when $Z$ increases from 0.02 to 0.1, see
Table~\ref{Tab:grids}). In those conditions, both $Z$ {\it and} $Y$ (and $X=1-Y-Z$)
determine the stellar properties as a function of metallicity.

  Chemical composition influences the stellar structure basically through three
contributions: the mean molecular weight $\mu$, the radiative opacity $\kappa$, and
the nuclear energy production $\varepsilon_{nuc}$. The metallicity-dependent
mass loss rate $\dot{M}$ further affects stellar evolution as a function of $Z$.

\subsection{The $\mu$-effect}
\label{Sect:mu-effect}

 The mean molecular weight acts in the hydrostatic equilibrium of the star.
Let us imagine a {\it thought
experiment} in which $\mu$ is arbitrarily increased over the whole star, all other
quantities such as $\kappa$ or $\varepsilon$ being unaltered (though keeping, of
course, their density and temperature dependences).  The reduction in the gas
pressure leads to an overall contraction (and heating) until hydrostatic equilibrium is re-established.
As a result, the stellar radius decreases and the surface temperature $T_{eff}$
increases. The surface luminosity $L$,
on the other hand, increases with $\mu$ since $L(r) \propto T^3\;\mathrm{d}T/\mathrm{d}r$
[in the case of radiative transfer, with $L(r)$ and $T$ being the luminosity and
temperature at radius $r$, respectively].
In summary, {\it an increase in $\mu$ tends to increase both $L$ and $T_{eff}$}.

 This $\mu$-effect is undetectable in low-metallicity stars, where $\mu$
is essentially constant as a function of metallicity. At metallicities higher
than about solar, however, $\mu$ increases as a result of increasing $Y$ [$Z$
remains unimportant in this respect for all realistic values of $Z$, since
$\mu \simeq (2 X + 0.75 Y + 0.5 Z)^{-1}$]. And indeed, it is shown in
Sections~\ref{Sect:MS} and \ref{Sect:CHeB} that very metal-rich models are more
luminous and hotter than solar-metallicity ones.
It has to be noted that
the $\Delta Y/\Delta Z$ law adopted in metal-rich models has an important
influence in this respect, as it fixes $Y$ and thus $\mu$.

\begin{table*}
\caption[]{\label{Tab:grids}
           Geneva group grids of models  (see text)
          }
\begin{tabular}{lccccll}
\hline
\noalign{\smallskip}
  Grid$^{(*)}$ & $Z$  &  $X$  &  $Y$  & $\dot{M}$ & Interior opacities & Low T opacities\\
\noalign{\smallskip}
\hline
\noalign{\smallskip}
 I   & 0.001 & 0.756 & 0.243 & standard       & Rogers \& Iglesias (1992) & Kurucz (1991) \\
 III & 0.004 & 0.744 & 0.252 & standard       & Iglesias et al. (1992)    & Kurucz (1991) \\
 II  & 0.008 & 0.728 & 0.264 & standard       & Rogers \& Iglesias (1992) & Kurucz (1991) \\
 I   & 0.020 & 0.680 & 0.300 & standard       & Rogers \& Iglesias (1992) & Kurucz (1991) \\
 IV  & 0.040 & 0.620 & 0.340 & standard       & Iglesias et al. (1992)    & Kurucz (1991) \\
 VII & 0.100 & 0.420 & 0.480 & standard       & Iglesias \& Rogers (1996) & Alexander \& Fergusson (1994) \\
 V   & $\!\!\!0.001\!\rightarrow\! 0.04$
             & \multicolumn{2}{c}{M$\geq$\mass{12}}
                             & 2$\times$standard & Iglesias et al. (1992) & Kurucz (1991) \\
\noalign{\smallskip}
\hline
\end{tabular}
\begin{tabbing}
 $^{(*)}$ \= III\= : Charbonnel et al. (1993, Paper III);\ \ \ \ \= VII\= \kill
 $^{(*)}$ \> I  \> : Schaller et al. (1992, Paper I);     \> IV  \> : Schaerer et al. (1993, Paper IV); \\
          \> II \> : Schaerer et al. (1993, Paper II);    \> V   \> : Meynet et al. (1994, Paper V);   \\
          \> III\> : Charbonnel et al. (1993, Paper III); \> VII \> : Mowlavi et al. (1997, Paper VII).\\
\end{tabbing}
\end{table*}

\subsection{The $\kappa$-effect}
\label{Sect:kappa-effect}

 The radiative opacity acts in the (radiative) transport of energy.
Let us consider again a {\it thought experiment} in which the opacity is
arbitrarily increased throughout the star, all other
quantities (such as $\mu$ or $\varepsilon_{nuc}$) being unaltered.
The hydrostatic structure remains unaffected to first approximation.
In particular the radius and temperature profiles are unchanged.
The energy flux, on the other hand, decreases with increasing $\kappa$ [since
$L(r) \propto \kappa^{-1}\;T^3\;\mathrm{d}T/\mathrm{d}r$ for radiative
transfer], leading to a decrease in the surface luminosity.
As the radius is unaffected to first order, the surface temperature
decreases too. In summary, {\it an increase in $\kappa$ tends to decrease both
$L$ and $T_{eff}$}.

 In general, the opacities increase with increasing metallicity.
This is the case for the bound-free (see Eq.~ 16.107 in Cox \& Giuli 1969)
and free-free (see Eq.~16.95 in Cox \& Giuli 1969) transitions.
In contrast, electron scattering depends only on $X$
[$\kappa_e\simeq 0.20(1+X)$], which is about constant at
$Z\simlr 0.01$ and decreases with $Z$ at $Z\simgr 0.01$.
As seen in Sect.~\ref{Sect:MS}, these considerations explain the stellar
properties as a function of $Z$ at sub-solar metallicities.

\subsection{The $\varepsilon_{nuc}$-effect}
\label{Sect:epsilon-effect}

 The nuclear energy production $\varepsilon_{nuc}$ sustains the stellar
luminosity. If $\varepsilon_{nuc}$ is arbitrarily increased, the central regions
of the star expand, leading to a decrease in the temperatures and densities, and
to an increase in the stellar radius. Thus, {\it an increase in $\varepsilon_{nuc}$
tends to decrease both $L$ and $T_{eff}$}, as well as the central temperature
$T_c$ and density $\rho_c$.

The way $\varepsilon_{nuc}$ depends on  metallicity is closely related to the
mode of nuclear burning.
When the CNO cycles are the main mode of H burning, $\varepsilon_{nuc}$ increases
with $Z$.
In contrast, when the main mode of burning is the pp chain, $\varepsilon_{nuc}$
is related to $X$. It is thus about independent of $Z$ at $Z\simlr 0.02$, and
decreases with increasing $Z$ at $Z\simgr 0.02$.
As shown in Sect.~\ref{Sect:MS}, these properties determine the behavior of
$T_c$ and $\rho_c$ in MS stars as a function of $Z$.

\subsection{The effect of $\dot{M}$}
\label{Sect:mass loss effect}

When mass loss is driven by radiation, Kudritzki et al. (1989) showed that $\dot{M}$
in O stars is proportional to $Z^{0.5}$. As a
result, the effects of mass loss dominate in more metal-rich stars. Mass
loss reduces the stellar mass, modifying thereby the post-MS
stellar properties (and even during the MS for the most massive ones).
It also turns the (massive) star into
a Wolf-Rayet star at an earlier stage than it would at lower metallicity. This is
analysed in more detail in Sect.~\ref{Sect:WR}.

\subsection{Summary}

At $Z\simlr 0.02$, $Z$ is the main factor acting through the $\kappa$-effect.
The adopted $\Delta Y/\Delta Z$ law is thus not influential.
At $Z\simgr 0.02$, on the other hand, $Y$
plays a dominant role mainly through the $\mu$-effect, while $Z$ acts
on the mass loss rate.  The adopted $\Delta Y/\Delta Z$ law is then important.

 These combined effects of $\mu$, $\kappa$ and $\varepsilon_{nuc}$ on the
stellar surface properties can be estimated more quantitatively with a
semi-analytical approach using homology relations. This is developed in the appendix for a
\mass{3} star.
A more detailed analysis of stellar properties with metallicity, however,
can only be done with the help of stellar model calculations. Such
analysis is provided
in Sects.~\ref{Sect:MS} and \ref{Sect:CHeB} using the Geneva grids of
models, whose ingredients are first summarized in the next section.

\section{Models}
\label{Sect:models}

  The stellar grids published by the Geneva group from $Z=0.001$ to 0.1
(summarized in Table~\ref{Tab:grids}) constitute a homogeneous set of models.
They are computed with
a value of $\alpha_p=l/H_p=1.6$ for the ratio of the mixing length $l$
in convective regions to the pressure scale height $H_p$ (determined in order to match
the sun
and the location of the red giant branch in the HR diagram), and
a moderate core overshooting distance of $d=0.20\;H_p$ beyond the classical
convective core boundary as defined by the Schwarzschild criterion (in order to
reproduce the MS width in the HR diagram). We refer to Paper I (see
Table~\ref{Tab:grids} for the references of the papers) for a general
description of the physical ingredients. We mention below only the specific
ingredients which differ from one set of calculations to another.

The primordial helium abundance
$Y_0$ is taken equal to 0.24 (cf. Audouze 1987) for all models.
The relative ratio of helium to metal enrichment, on the
other hand, is $\Delta Y/\Delta Z=3$ at $Z\leq 0.02$, 2.5 at $Z=0.04$, and
2.4 at $Z=0.1$. At low metallicity, the model output is not very sensitive to
the adopted value for $\Delta Y/\Delta Z$. At $Z\simgr 0.02$, however, the model
output is sensitive to $\Delta Y/\Delta Z$ (see Sect.~\ref{Sect:chemical
composition}). The values adopted at $Z=0.04$ and 0.1 are consistent with the
currently admitted value of $\sim 2.5$ (see Peimbert 1995 and references
therein), but the uncertainties are still very high (Maeder 1998).

 The grids presented in Papers I to VII are all calculated with the OPAL
opacities. They use, however, different OPAL tables according to their
availability at the time of each calculation, which are
summarized in Table~\ref{Tab:grids}. The low
temperature opacities are taken from Kurucz (1991) in all grids, except
in Paper VII which uses the Alexander \& Ferguson (1994) tables.

  The mass loss prescription is of crucial importance for the evolution of the
models. A general dependence of $\dot{M}\propto Z^{0.5}$, as suggested by the
models of Kudritzki et al. (1989),
is adopted for all the stars except the WR stars. 
A Reimers law (1975) is adopted in all grids for low- and
intermediate-mass stars. For massive stars, the situation is more delicate. In a
first set of calculations (Papers I to IV), mass loss rates from de Jager et al.
(1988, `standard' mass loss rates) were used. A second set of calculations (Paper
V) were then performed with enhanced mass loss rates during the pre-WR and WNL
phases (by a factor of two over the `standard' ones). The latter models were found
to better reproduce many observational constraints concerning massive stars
(Maeder \& Meynet 1994), and we use them in this paper too. At $Z=0.1$, however,
the standard mass loss rates for massive stars are used (Paper VII). Indeed, we
expect the mass loss rates driven by radiation pressure to be probably saturated
at such high metallicities. The conclusions are, however, qualitatively not
affected by this choice.

Let us recall that at $Z=0.1$, the criterion defining a star to be of the
Wolf-Rayet type could be different from that at lower metallicities. However,
in the
absence of any observational counterpart of those objects at such high
metallicities, we keep the same criterion as in Paper I, i.e. we consider that
a star enters the WR stage when its surface hydrogen mass fraction becomes lower
than 0.4 and its effective temperature is higher than 10000~K. These values are
actually not crucial for the analysis of WR populations since the characteristic
time-scale for entering the WR stage is very short compared to the WR lifetimes.
The mass loss rates of these objects are then accordingly taken as in Paper I.

\section{The main sequence}
\label{Sect:MS}

\subsection{The zero-age main sequence}
\label{Sect:ZAMS}

\begin{figure}
  \resizebox{\hsize}{!}{\includegraphics{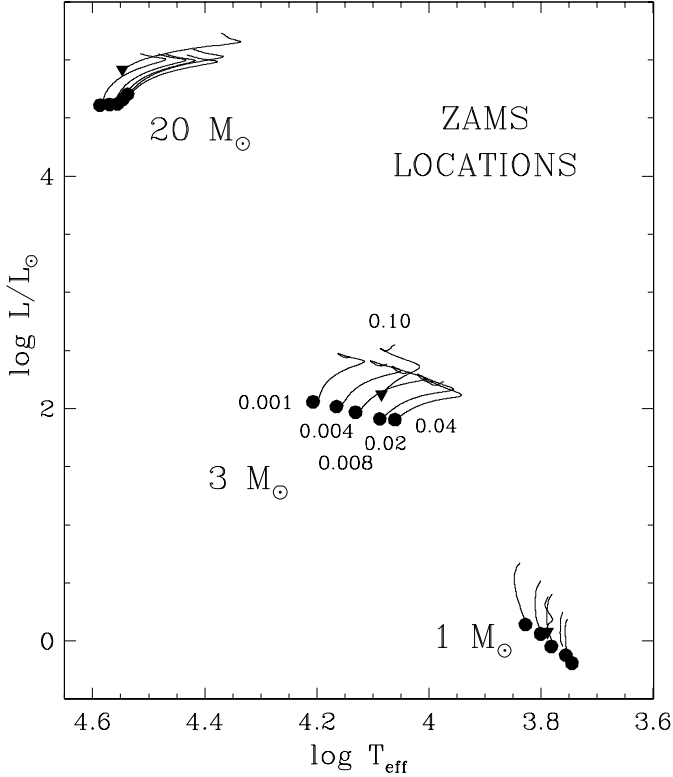}}
  \caption{ZAMS locations in the HR diagram of 1, 3 and \mass{20} star
           models at different metallicities as labeled in the figure for the
           \mass{3} star. Models at $Z=0.1$ are identified by triangles. The
           models are computed with OPAL and Alexander \& Ferguson (1994)
           opacities}
  \label{Fig:HR}
\end{figure}

\begin{figure}
  \resizebox{\hsize}{!}{\includegraphics{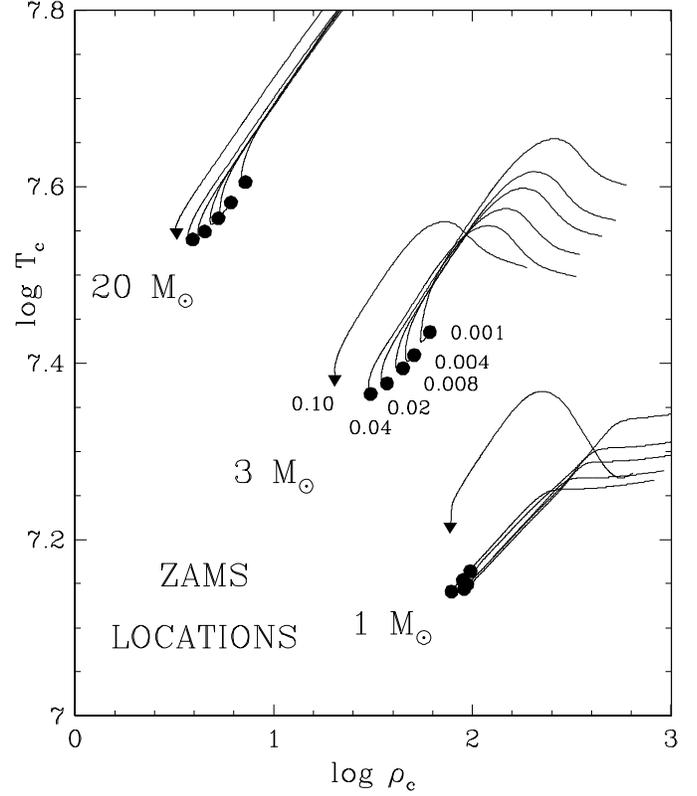}}
  \caption{Same as Fig.~\ref{Fig:HR} but in the ($\log \rho_c, \log T_c$) diagram}
  \label{Fig:roT}
\end{figure}

  The arguments presented in Sect.~\ref{Sect:chemical composition} enable to
understand the location of the zero-age main sequence (ZAMS) in the 
Hertzsprung-Russell (HR) diagram as a function of metallicity. That location is
given in Fig.~\ref{Fig:HR} for stars of 1, 3 and \mass{20} at metallicities
ranging from $Z=0.001$ to 0.1.

\paragraph{Metallicities below $\simeq 0.04$:}

 At $Z\simlr 0,.4$, the surface
properties (as a function of $Z$) are determined by the $\kappa$-effect (see
Sect.~\ref{Sect:kappa-effect}). The surface temperature is mainly
sensitive to the opacity in the outer layers where bound-bound and bound-free
transitions dominate, so that all ZAMS models
get cooler with increasing metallicities. The surface luminosity, on the other
hand, is sensitive to the interior opacities. In low- and intermediate-mass
stars, the main source of opacity is provided by bound-free transitions, and
increases with metallicity. As a result, $L$ decreases with $Z$. In massive
stars, however, electron-scattering is the main source of opacity. As this source
is metallicity independent, the luminosity of low-metallicity massive stars is
insensitive to $Z$.

\paragraph{Metal-rich stars:}

As the metallicity increases above $Z\simeq 0.02$, the $\mu$-effect becomes
competitive with the $\kappa$-effect. In low- and intermediate-mass stars,
the $\mu$-effect eventually exceeds the $\kappa$-effect above $Z\simeq 0.05$,
and a further increase in the metallicity results in {\it hotter} and
{\it more luminous}\footnote{These
properties remained unnoticed by Fagotto et al. (1994), though inspection of the
data available from their tables reveal that their results indeed support our
conclusions.}.
ZAMS models (see Sect.~\ref{Sect:mu-effect}).
In massive stars, $L$ already begins to increase as a
function of $Z$ at $Z\simeq 0.02$ since the $\kappa$-effect on $L$ is negligible
in those stars.

Let us now consider the central temperature $T_c$ and density $\rho_c$.
Those quantities are, contrarily to $L$ and $T_{eff}$, closely related to the
nuclear energetic properties of the core material (see
Sect.~\ref{Sect:epsilon-effect}).
In particular, they are sensitive to the mode of hydrogen
burning. Usually, the CNO cycles are the main mode of core H burning
for stars more massive than about \mass{1.15-1.30} (depending on metallicity),
while the pp chain operates at lower masses. At $Z=0.1$, however, all stars
are found to burn their hydrogen through the CNO cycles (at least down to
$M=\mmass{0.8}$ considered in our grids).

 When the CNO cycle is the main mode of burning, $\rho_c$ and $T_c$ decrease with
increasing $Z$ (see Sect.~\ref{Sect:epsilon-effect}).
This is clearly visible
in Fig.~\ref{Fig:roT} for the 3 and \mass{20} stars at $Z\leq 0.04$.
At $Z=0.1$, however, the higher surface luminosities of the models as compared to
those at $Z=0.04$ result in concomitant higher central temperatures.

 When the pp chain is the main mode of burning, on the other hand,
the location in the $(\log \rho_c,\log T_c)$ is not very sensitive to $Z$.
This is well verified for the \mass{1} models at $Z<0.1$
(Fig.~\ref{Fig:roT}).

\subsection{Main sequence convective core sizes}
\label{Sect:ZAMSQcc}

\begin{figure}
  \resizebox{\hsize}{!}{\includegraphics{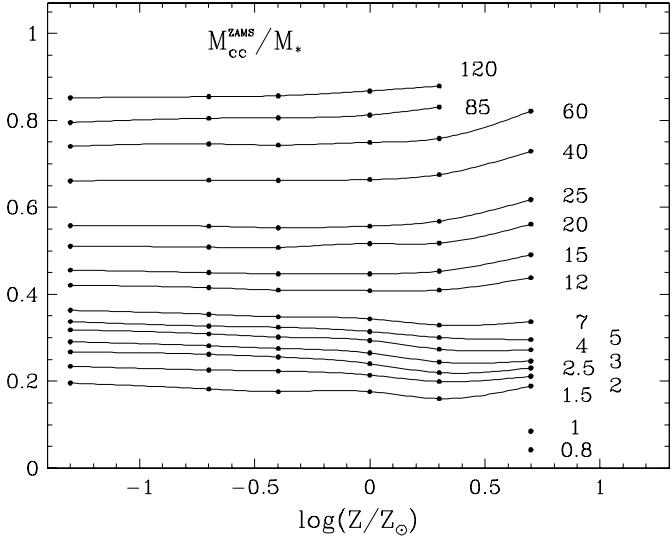}}
  \caption{Masses of convective cores relative to the stellar masses
           as labeled (in solar masses) next to the curves, as a function
           of metallicity}
  \label{Fig:ZAMSQcc}
\end{figure}

  The mass of the convective core is mainly determined by the nuclear energy production,
and thus by $L$.  At $Z\simlr 0.04$, it
slightly decreases with increasing $Z$ in low- and intermediate-mass stars, while it
remains approximately constant in massive stars (see Fig.~\ref{Fig:ZAMSQcc}).
At $Z\simgr 0.04$, on the other hand, it increases with $Z$ at all stellar
masses\footnote{It is interesting to note that we could have expected smaller
convective cores in metal-rich massive stars where the main source of opacity is
electron scattering (since this opacity is positively correlated with the H content). But
the effect of higher luminosity in those stars overcomes that $\kappa$-effect, and
the core mass increases with increasing $Z$.}.
We recall furthermore that at $Z=0.1$ all stars with $M$ as low as \mass{0.8}
possess convective cores.

\subsection{Main sequence lifetimes}
\label{Sect:tH}

\begin{figure}
  \resizebox{\hsize}{!}{\includegraphics{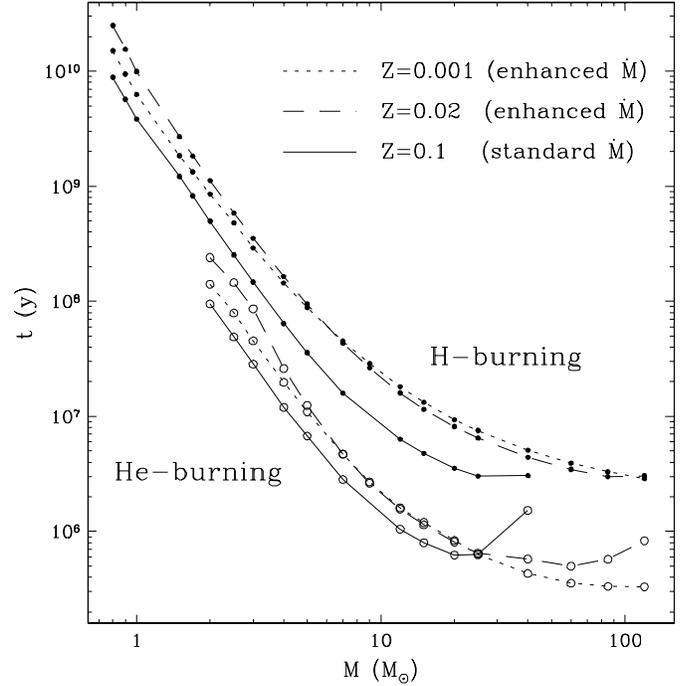}}
  \caption{Main sequence (filled circles) and core helium burning (open circles)
           lifetimes of Z=0.001 (dotted lines), 0.02 (dashed lines) and 0.1 (solid lines)
           models as a function of their initial mass}
  \label{Fig:t}
\end{figure}

\begin{figure}
  \resizebox{\hsize}{!}{\includegraphics{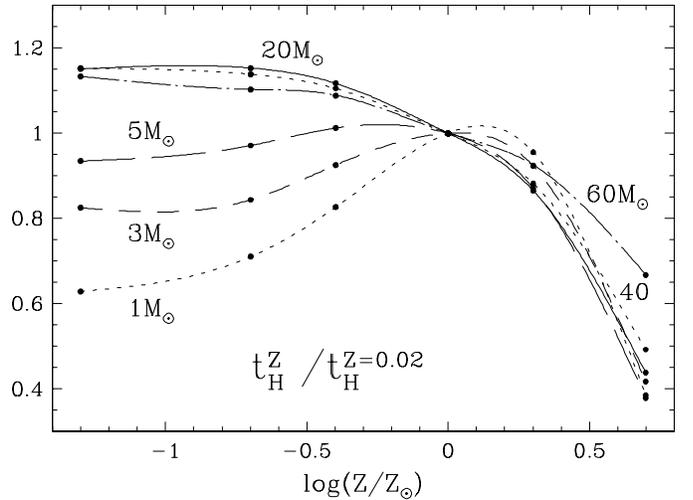}}
  \caption{Main sequence lifetimes of models of initial masses as labeled on
           the curves, as a function
           of their metallicity. The lifetimes are normalized for each stellar
           mass to its value at $Z=0.02$}
  \label{Fig:tH}
\end{figure}

  The MS lifetimes $t_H$ are summarized in Fig.~\ref{Fig:t} for three different
metallicities as a function of the initial stellar mass, and
in Fig.~\ref{Fig:tH} for several stellar masses as a function of metallicity.
They can be understood in terms
of two factors: the initial H abundance, which determines the quantity of
available fuel, and the luminosity of the star, which fixes the rate at which
this fuel burns.

At $Z\simlr 0.02$, $t_H$ is mainly determined by $L$, being shorter at higher
luminosities. Fig.~\ref{Fig:tH} confirms, as expected from $L$ (see
Fig.~\ref{Fig:HR}), that $t_H$
increases with $Z$ in low- and intermediate-mass stars, and is about independent
of $Z$ in massive stars.

At $Z > 0.02$, on the other hand, the initial H abundance decreases sharply with
increasing $Z$ (the H depletion law is dictated by the adopted $\Delta Y/\Delta
Z$ law; for $\Delta Y/\Delta Z=2.4$ and $Y_0=0.24$, $X$ drops from 0.69 at $Z=0.02$
to 0.42 at $Z=0.1$). The MS
lifetimes are then mainly dictated by the amount of fuel available.
This, combined with the higher luminosities at $Z=0.1$, leads
to MS lifetimes which are about 60\% shorter at $Z=0.1$ than at
$Z=0.02$. This result is independent of the stellar mass for $M\leq \mmass{40}$. Above
this mass, however, the action of mass loss extends the MS lifetime,
as can be seen from the \mass{60} curve in Fig.~\ref{Fig:tH}.

\subsection{Masses at the end of the MS phase}
\label{Sect:MH}

\begin{figure}
  \resizebox{\hsize}{!}{\includegraphics{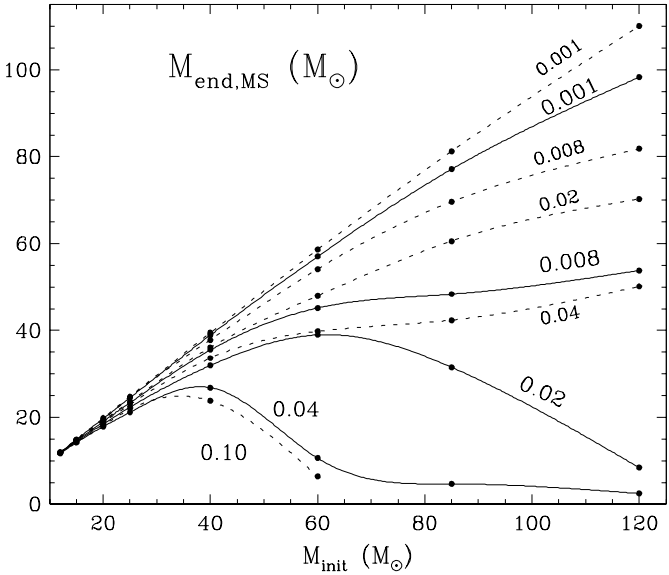}}
  \caption{Stellar masses at the end of the main sequence phase as a function
           of initial mass for different metallicities as labeled on the curves.
           Dotted lines correspond to models computed with the `standard' mass
           loss rates for massive stars (Papers~I to IV and VII), while thick lines
           correspond to models computed with twice that `standard'
           mass loss rates (Paper~V)}
  \label{Fig:MH}
\end{figure}

  Mass loss determines crucially the evolution of massive stars, especially at
high metallicities.  The stellar masses remaining at the end of the
MS phase with the `enhanced' mass-loss prescription (see Sect.~\ref{Sect:models})
are shown in thick lines in Fig.~\ref{Fig:MH} (the ones with `standard' mass-loss
prescription is also shown in dotted lines for comparison).
The most striking result at metallicities higher than twice solar
is the rapid evaporation of massive stars with
$M\simgr\mmass{50}$ , and the
consequent formation of WR stars during core H burning.
At $Z=0.04$, stars more
massive than $\sim$\mass{80} leave the core hydrogen burning  phase
with less than \mass{5}. The same
holds true at $Z=0.1$ even though `standard' mass loss rates are
used for that metallicity.

\section{The core He-burning phase}
\label{Sect:CHeB}

\begin{figure}
  \resizebox{\hsize}{!}{\includegraphics{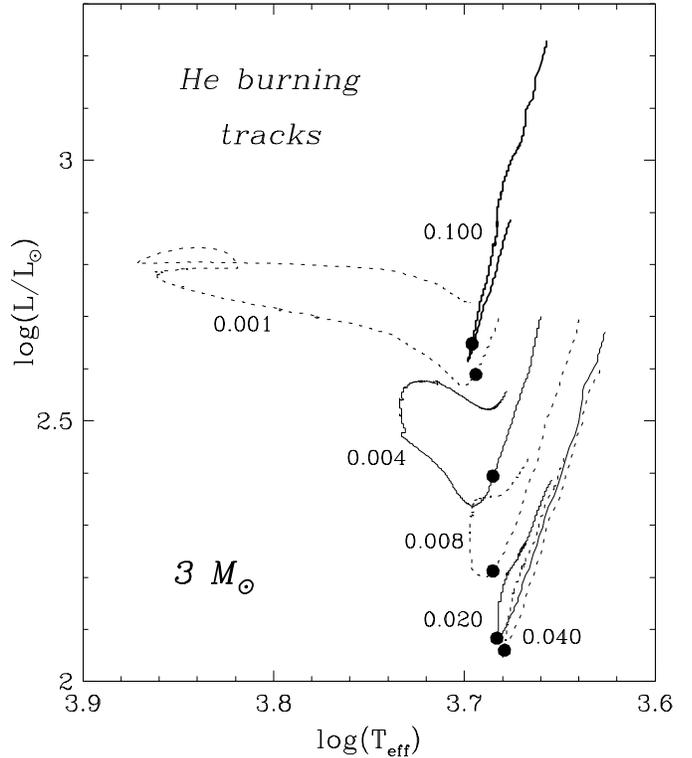}}
  \caption{Tracks followed in the HR diagram during the core helium burning phase by
           \mass{3} models at different metallicities as labeled in the figure.
           The points on each track locate the model when the core helium
           abundance has dropped to $X_c({\rm ^4He})=0.85$}
  \label{Fig:HeBhr3}
\end{figure}

\begin{figure}
  \resizebox{\hsize}{!}{\includegraphics{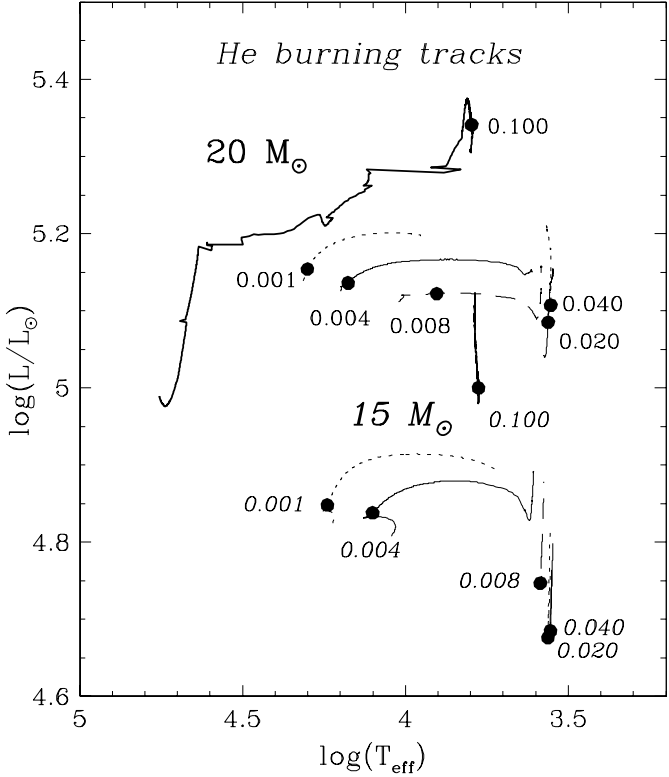}}
  \caption{Same as Fig.~\ref{Fig:HeBhr3} but for the \mass{15} (labeled in
           italics) and \mass{20} (labeled in straight) models. Only the part
           following He ignition is shown}
  \label{Fig:HeBhr20}
\end{figure}

\subsection{The HR diagram}

The surface properties during the CHeB phase are dictated by the same arguments
developed in Sect.~\ref{Sect:chemical composition} and
used to explain the MS properties (see
Sect.~\ref{Sect:MS}). At $Z\simlr 0.04$, both $L$ and $T_{eff}$ decrease with
increasing $Z$, due to the $\kappa$-effect. At
$Z\simgr 0.04$, on the other hand, both $L$ and $T_{eff}$ increase with $Z$, due
to the $\mu$-effect. These properties
are well verified in Figs.~\ref{Fig:HeBhr3} and \ref{Fig:HeBhr20} which show the
CHeB tracks in the HR diagram of intermediate-mass and massive stars,
respectively.

  Low- and intermediate-mass stars of low metallicity are known to perform a
blue loop in the HR diagram and cross the Cepheid instability strip.
The extent of that loop is a function of both the stellar mass (more massive stars having
more extended loops) and the metallicity (more metal-rich stars having
less extended blue loops). For a \mass{3} star, the blue loop is suppressed at
$Z\simgr 0.008$ (see Fig.~\ref{Fig:HeBhr3}), while it is suppressed at $Z>0.02$ for
a \mass{5} star. Furthermore, no blue loop is obtained for any stellar mass at
metallicities higher than $Z\simeq 0.04$.

  Another interesting property is the effective temperature
at which massive stars enter the CHeB phase. Low-metallicity massive star
models ignite helium as blue supergiants, while
solar metallicity models ignite helium as red supergiants.
This location of helium ignition as a function of $Z$
in the HR diagram is shown in Fig.~\ref{Fig:HeBhr20}.
As a result, the number ratio of blue-to-red supergiants is a decreasing function
of the metallicity\footnote{Let
us recall that this property is found by all existing stellar models, but is {\it
not} supported by the observations.  We refer to Langer \& Maeder (1995) for a
discussion on this issue.}.
We finally notice, again, the departure of $Z=0.1$ models to the general rule: they
ignite helium at a surface temperature higher than those of solar metallicity
models. This is due to the $\mu$-effect (see Sect.~\ref{Sect:mu-effect}).

\subsection{Core helium burning lifetime}

\begin{figure}
  \resizebox{\hsize}{!}{\includegraphics{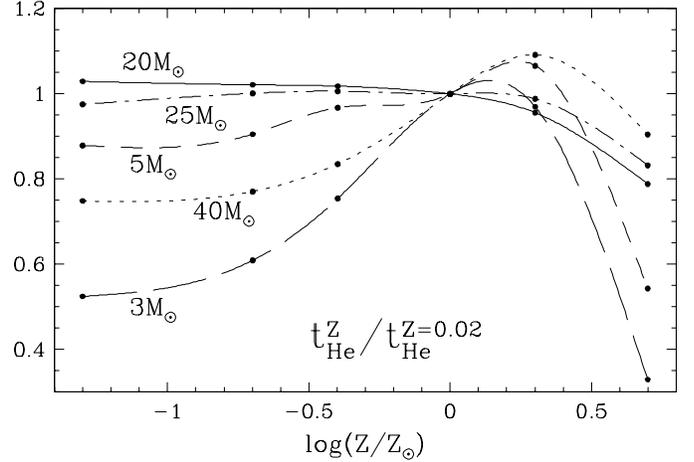}}
  \caption{Core He burning lifetimes of models of initial masses as labeled on
           the curves, as a function
           of metallicity. The lifetimes are normalized for each stellar
           mass to its value at $Z=0.02$}
  \label{Fig:tHe}
\end{figure}

  The CHeB lifetime $t_{He}$ is displayed in Fig.~\ref{Fig:t} for three different
metallicities as a function of the initial stellar mass, and in
Fig.~\ref{Fig:tHe} as a function of
metallicity for several initial masses.  It follows a pattern very similar
to that of $t_H$ (Fig.~\ref{Fig:tH}) for intermediate-mass stars, and can be
understood on grounds of the same arguments developed in Sect.~\ref{Sect:tH}.

  In massive stars, the CHeB lifetime is extended as a result of the lower average
luminosities due to mass
loss.  As a result, the CHeB lifetime relative to its value at $Z=0.02$ increases
with the initial stellar mass.

\begin{figure}
  \resizebox{\hsize}{!}{\includegraphics{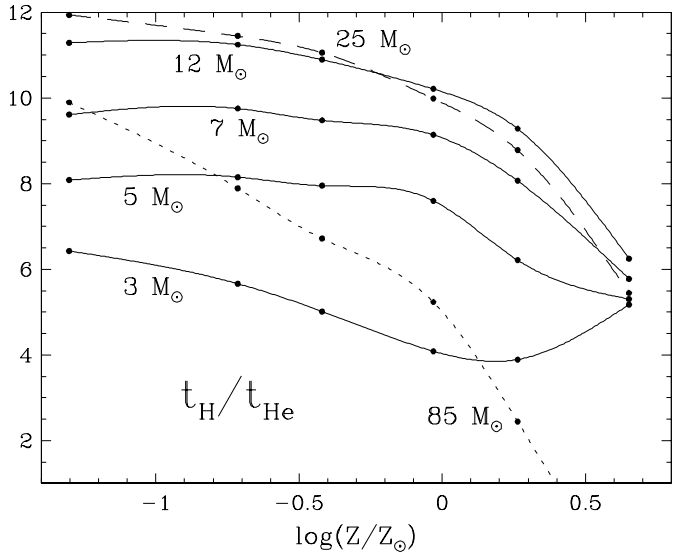}}
  \caption{Ratio of core H to core He burning lifetimes for models of
           initial masses as labeled on the curves, as a function of metallicity}
  \label{Fig:tHetH}
\end{figure}

 The ratio $t_H/t_{He}$ is displayed in Fig.~\ref{Fig:tHetH}. Below $Z\simeq 0.04$,
this ratio is a decreasing function of metallicity, for a given stellar mass, and
an increasing function of stellar mass, for a given metallicity. Above $Z\simeq
0.04$, this ratio converges to the value of $t_{H}/t_{He}\simeq 5.5$, irrespective of
the stellar mass. For stars more massive than $\sim$\mass{30}, however, the
high mass loss rates reduce further this ratio.

\subsection{Masses at the end of the CHeB phase}
\label{Sect:MHe}

\begin{figure}
  \resizebox{\hsize}{!}{\includegraphics{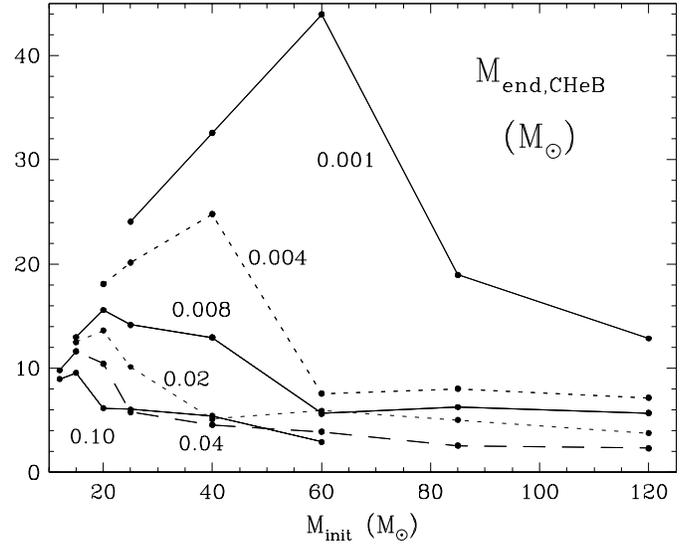}}
  \caption{Stellar masses at the end of the core helium burning phase as a
           function of initial mass for different metallicities as labeled on
           the curves. Only models computed with twice the
           `standard' mass loss rates for massive stars are shown, except at
           $Z=0.1$ for which the mass loss rates are 'standard'}
  \label{Fig:MHe}
\end{figure}

  The masses at the end of the CHeB phase of massive stars are displayed in
Fig.~\ref{Fig:MHe} for several metallicities. These are a decreasing function
of increasing metallicity.
%It is also a decreasing function of $M$ for stars less massive than
%about \mass{40} at $Z\ge 0.02$, and \mass{60} at $Z=0.008$ and $Z=0.004$.

\section{Wolf-Rayet stars in very metal-rich environments}
\label{Sect:WR}

The properties of the WR star models at metallicities between 0.001 and 0.04
have already been presented in Maeder \& Meynet (1994). Here we shall focus our
discussion on
the expected characteristics of the WR stars in very metal-rich environments,
i.e. at $Z= 0.1$. 

The minimum initial mass for a single star to become a WR star is a decreasing
function of metallicity. It goes from \mass{25} at $Z=0.02$
to \mass{21} at $Z=0.04$ (see table 1 in Maeder \& Meynet 1994, $2\times\dot{M}$
case), and to approximately \mass{17} at $Z=0.1$.

\subsection{$M_{init}\simgr\mmass{60}$: ending as helium white dwarf?}

At $Z=0.1$, stars more massive than about 60 $M_\odot$ literally evaporate
during the main sequence phase. They loose so much mass that they never reach
the He-burning phase, and will likely end their nuclear life as helium white dwarfs.
Let us remark here that helium white dwarf stars are generally thought to be
the end point evolution of very low mass stars typically with initial masses
between 0.08 and 0.5 $M_\odot$. Their existence in our present universe is thus
doubtful (unless created through binary mass transfer), since the MS lifetimes of
these very low mass stars are greater than the present age of the universe.
In contrast, the above scenario (already addressed by Maeder \& Lequeux 1982; see
also Paper V), in which massive evaporating stars could end their stellar
life as helium white dwarfs, allows the formation of such objects in 
young star forming regions at very high metallicities.

\subsection{WN {\rm \&} WC Wolf-Rayet stars}

Stars more massive than about 50 $M_\odot$ never enter the WC
phase at $Z=0.1$. Due to very high mass loss rates acting already during the MS, the
H-convective core rapidly decreases in mass, leaving behind a great portion
of the star with CNO-enriched material. As a consequence, the star goes
through a very long WN phase. Moreover, these stars enter the core
He-burning phase, if they do enter this phase at all, with very low masses. This
implies in particular so small mass loss rates, according to the
$\rm \dot{M}(M)$ relation proposed by Langer (1989), that their helium
burning core never uncovers, preventing the star to become a WC Wolf-Rayet star. 

\subsection{Wolf-Rayet versus O type stars}
\label{Sect:WR-O}

The number ratio of WR to O type stars expected in a constant star formation
rate region at $Z=0.1$ is estimated from these models to be around 0.45.
To compute this value, only the single star evolutionary channel was
considered, we chose a Salpeter initial mass function (i.e. with a slope equal
to 1.35), and the upper mass limit was taken equal to 60 $M_\odot$.

From the above result, one expects to find at $Z=0.1$ about
1 WR star per each 2 O-type stars detected! This is a very large proportion
compared to regions with solar and sub-solar metallicity
(in the solar neighborhood, the observed number ratio WR/O is 0.1).
This high proportion results mainly from the very high mass losses experienced
by these metal-rich stars, which, as seen above, decrease the minimum initial
mass for WR star formation, make the star enter the WR phase at an earlier
stage of its evolution, and increase the core helium burning lifetime. 
Moreover, the duration of the O-type phase of metal-rich stars is considerably
reduced as compared to that at lower metallicities. For instance when Z increases
from 0.04 to 0.1, the lifetime of the O-type phase decreases by about a factor 2. This
is essentially a consequence of the shorter duration of the H-burning phase at $Z=0.1$
(see Fig.~\ref{Fig:tH}).

\subsection{Supernovae}

Let us now consider the final evolution of massive stars and estimate
the relative fraction of supernovae with WR progenitors using some
simplifying assumptions (see also Meynet et al., 1998).
There is at present some uncertainties concerning the fate of WR stars, and to
which type of supernovae they can give birth. As a first approximation, we make
the hypothesis that all WR stars do lead to supernovae events (see a discussion
in Maeder \& Lequeux 1982). Moreover, since
the H-rich envelope is ejected through the wind during the WNL phase, we further
assume that WNE and WC stars result in supernovae of type I (whether Ib, Ic, or
another yet unknown subclassification).
With these assumptions, and not considering type Ia
supernovae which have low-mass progenitors,
one expects that metal-rich regions be characterized by a higher fraction of
type I(b/c) supernovae relative to the total number of supernovae (Maeder 1992).
  As a numerical example, using
the same IMF as in Sect.~\ref{Sect:WR-O} and making the hypothesis of constant
star formation rate, one obtains that, at $Z=0.1$, the fraction of
supernovae with WNE-WC progenitors amounts to 32\% of the total number of supernovae having
massive star progenitors. In other words, 1 supernova event out of 3
should be of type I(b/c) in very metal-rich regions.  For comparison, this
proportion decreases to 1 WR explosion per $5 - 7$ supernovae at solar
metallicity.

\section{Conclusions}
\label{Sect:conclusions}

  This paper analyzes some properties of stellar models as a function of the
metallicity, with particular attention to those at very high metallicities
(above twice or three times solar). The stellar properties as a function of $Z$
are shown to result from the action of
four ingredients\footnote{We neglect in this paper the effects of mild mixing
processes due for example to rotation, diffusion, gravitational settling, etc...
which may modify the internal composition profiles and some of the values given
here}:
the mean molecular weight, an increase of which leads to more
luminous and hotter stars (the $\mu$-effect);
the opacity, an increase of which leads to less
luminous and cooler stars (the $\kappa$-effect);
the nuclear energy production (but to a lesser
extent), an increase of which leads to less luminous and cooler stars with lower
$T_c$ and $\rho_c$ (the $\varepsilon_{nuc}$-effect;
and mass loss, which acts indirectly on the stellar structure by reducing
the stellar mass (dramatically in the most massive
metal-rich stars).
The $\kappa$-effect plays the main role at $Z\simlr 0.04$, while the
$\mu$-effect becomes preponderant at $Z\simgr 0.02$.
In the later case, the $\Delta Y/\Delta Z$ law is important.

  Based on these considerations and on the Geneva's grids of stellar models, it
is shown in particular that very metal-rich stars exhibit properties which cannot
be extrapolated from model characteristics at lower metallicities. In
particular,

\noindent - very metal-rich stars (such as at $Z=0.1$) are more luminous and
hotter than those at $Z\le 0.05$.
This property contrasts with the usually known trend at lower metallicities (whereby
$L$ and $T_{eff}$ decrease with increasing $Z$ due to the $\kappa$-effect);

\noindent - very metal-rich stars have much lower MS lifetimes than those at
$Z\le 0.05$. For example the MS lifetime at $Z=0.1$ is 2.5 times shorter than
that at solar metallicity, independent of their initial mass (at least up to
\mass{40});

\noindent - stellar population synthesis in metal-rich environments is
significantly affected by the increased mass loss rates at high $Z$.
In particular,
a) massive stars with $Z=0.1$ enter the WNL Wolf-Rayet phase very early
during their evolution, and those more massive than about \mass{50} probably never
enter the WC phase;
b) the most massive ($M\ge\mmass{60}$ at $Z=0.1$) high-metallicity stars loose the
majority of their mass during their MS phase, and probably end their life as He
white dwarfs;
and c) the number of WR stars over O-type stars increases with metallicity, 
and metal-rich regions are expected to be characterized by a high fraction of
type I (b/c/other?) supernovae relative to the total number of supernovae having massive star
progenitors (up to 1 out of 3 core collapse supernovae events at $Z=0.1$).

\appendix
\section*{Appendix: The ZAMS location in the HR diagram as a function of
metallicity: a semi-analytical approach}
\label{appendix}

The luminosity and effective temperature on the zero age main sequence (ZAMS)
are
given by the homology relations (see, for example, Cox and Giuli 1969, p~696)
\begin{equation}
  \label{Eq:L}
  L \propto \epsilon_0^{\ontops{-0.02}{-0.08}}
            \kappa_0^{\ontops{-1.02}{-1.08}}
            \mu^{\ontops{7.3}{7.8}}
            M^{\ontops{5.2}{5.5}}
\end{equation}
and
\begin{equation}
  \label{Eq:Teff}
  T_{eff} \propto \epsilon_0^{\ontops{-0.03}{-0.10}}
                  \kappa_0^{\ontops{-0.28}{-0.35}}
                  \mu^{\ontops{1.6}{2.2}}
                  M^{\ontops{0.94}{1.33}}
\end{equation}
where $\mu$ and $M$ are the mean
molecular weight and the stellar mass, respectively, $\epsilon_0$ the
temperature
and density independent coefficient in the relation for the nuclear energy
production $\epsilon_{nuc}=\epsilon_0\;\rho^\lambda\;T^\nu$ (see Cox and Giuli
1969, p~692), and $\kappa_0$ the
opacity coefficient in the Kramers law $\kappa=\kappa_0\;\rho\;T^{-3.5}$.
The numbers on the first line of the superscripts in Eqs.~\ref{Eq:L} and
\ref{Eq:Teff}
apply when the nuclear energy production results from the $pp$ chain, while the
second
line applies when the $CNO$ cycles provide the main source of nuclear energy.

A variation $\Delta Z$ in the metallicity of a star of mass $M$ affects
its position in the HR diagram by
\begin{eqnarray}
  \label{Eq:DeltaL0}
  \lefteqn{\Delta \log L =   \left\{\ontop{7.3}{7.8}\right\} \Delta \log \mu
                  - \left\{\ontop{1.02}{1.08}\right\} \Delta \log \kappa_0}
\hspace{4.5cm} \nonumber \\
  & &             - \left\{\ontop{0.02}{0.08}\right\} \Delta \log \epsilon_0
\end{eqnarray}
\begin{eqnarray}
  \label{Eq:DeltaTeff0}
  \lefteqn{\Delta \log T_{eff} =   \left\{\ontop{1.6}{2.2}\right\} \Delta \log
\mu
                        - \left\{\ontop{0.28}{0.35}\right\} \Delta \log
\kappa_0} \hspace{4.5cm} \nonumber \\
  & &                   - \left\{\ontop{0.03}{0.10}\right\} \Delta \log
\epsilon_0
\end{eqnarray}

The variation of $\mu$ is easily related to that of $Z$:
\begin{equation}
  \label{Eq:Deltamu}
  \Delta \log \mu= \frac{4.5}{\ln 10} \mu \Delta Z
\end{equation}
where we have used the relations
\begin{equation}
  \label{Eq:YZ}
  Y=0.24+2.4Z
\end{equation}
and
\begin{equation}
  \label{Eq:XYZ}
  X+Y+Z=1.
\end{equation}

\begin{figure}
  \resizebox{\hsize}{!}{\includegraphics{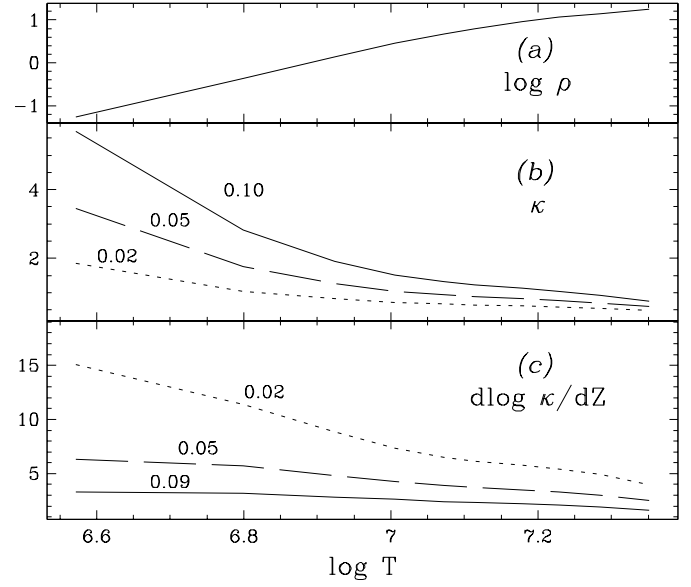}}
  \caption{{\sl(a)}: Density profile as a function of temperature in a ZAMS model
                     of a \mass{3} stars at $Z=0.1$;
           {\sl(b)}: Opacity profiles, as a function of temperature, at three
                     metallicities as labeled on the curves, and calculated
                     from the tables of Iglesias \& Rogers (1996) at the densities
                     displayed in {\sl(a)};
           {\sl(c)}: Same as {\sl(b)}, but for the derivative
                     $\mathrm{d}\log \kappa/\mathrm{d}Z$ at
                     the labeled metallicities. The derivative is calculated
                     numerically from
                     $(\log \kappa^{Z+0.01} - \log \kappa^{Z-0.01})/0.02$}
  \label{Fig:kappa}
\end{figure}

The variation of $\Delta \log \kappa_0$ as a factor of $\Delta Z$ is more
difficult to evaluate. We can get an estimation of
$\Delta \log \kappa_0 / \Delta Z$ by comparing the opacities given by the
Rogers and Iglesias (1992) tables at different metallicities.
The opacities at $Z=0.02$, 0.06 and 0.1 (and at $X, Y$ given by Eqs.~\ref{Eq:YZ}
and \ref{Eq:XYZ}) are shown in Fig.~\ref{Fig:kappa}{\sl(b)} for temperature and
density conditions representative of the interior of a \mass{3}, $Z=0.1$ ZAMS model
[Fig.~\ref{Fig:kappa}{\sl(a)}; we assume, as a first approximation, that the
same $\rho-T$ profile applies at those three metallicities]. If the opacity profile
were following a Kramers law, then the derivative
$\mathrm{d}\log \kappa / \mathrm{d}Z$ would be constant and equal to that of
the Kramers opacity coefficient $\mathrm{d}\log \kappa_0 / \mathrm{d}Z$ ($\rho$
and $T$ being held constant). Inspection of Fig.~\ref{Fig:kappa}{\sl(c)} shows
that this is obviously not the case.
%The temperature ranges from about $3.7\times 10^6$ to
%$22.4\times 10^6$~K, and covers 98\% of the total mass of the star.
%We apply the homology relation \ref{Eq:L} to that portion of the
%star comprising \mass{2.94}, and assume that the
%luminosity is constant in the remaining 2\% of the stellar mass.
%The $\Delta \log \kappa / \Delta Z$ ratio can now be evaluated numerically
%at a given metallicity $Z$ by calculating $(\log \kappa^{Z+0.005} -
%\log \kappa^{Z-0.005})/0.01$. The value of this numerical derivative is plotted
%in Fig.~\ref{Fig:kappa}{\sl(c)} for several metallicities.
The derivative is seen to vary
from 4 to 15 at $Z=0.02$ and from 1.5 to 3.3 at $Z=0.09$. However in this order of
magnitude approach, we consider mean values equal
\begin{eqnarray}
  \label{Eq:DeltaZ02}
  \Delta \log \kappa_0 \simeq 10 \Delta Z & \hspace{2cm} & {\rm Z=0.02}\\
  \label{Eq:DeltaZ10}
  \Delta \log \kappa_0 \simeq 2.5 \Delta Z & & {\rm Z=0.1}
\end{eqnarray}
%but keep in mind the significant departure of the opacity from a Kramers law,
%limiting thereby the quantitative capability of our semi-analytical approach.

The variation of $\Delta \log \epsilon_0$ as a function of heavy element
abundances $\Delta Z$ can be estimated knowing that the CN cycle is the main
energy producer in the \mass{3} star (see e.g. Mowlavi 1995, Fig.~A.3). The two
reactions \reac{C}{12}{p}{\gamma}{N}{13} and \reac{C}{13}{p}{\gamma}{N}{14}
contribute for more than 70\% of the total nuclear energy at the center of the
star. We can thus write that $\epsilon_0 \propto Z \times X$, which leads to,
using
Eqs.~\ref{Eq:YZ} and \ref{Eq:XYZ},
\begin{equation}
  \label{Eq:Deltaepsilon}
  \Delta \log \epsilon_0 = \left( \frac{1}{Z}-\frac{3.4}{X} \right)
\frac{1}{\ln 10} \Delta Z
\end{equation}

Equations \ref{Eq:DeltaL0} and \ref{Eq:DeltaTeff0} become with
Eqs.~\ref{Eq:Deltamu},
\ref{Eq:DeltaZ02}, \ref{Eq:DeltaZ10} and \ref{Eq:Deltaepsilon}, and taking
$X=0.70$, $\mu=0.617$ for $Z=0.02$ and $X=0.42$, $\mu=0.80$ for $Z=0.1$,
\begin{eqnarray}
  \label{Eq:DeltaLZ}
  \frac{\Delta \log L}{\Delta \log Z} =
       \left\{ \begin{array}{cccccccr}
                 0.41&\!\!\!-\!\!\!&0.50&\!\!\!-\!\!\!&0.07&\!=\!&-0.16
&\;\;\;\;\;\;Z=0.02\\
                 2.81&\!\!\!-\!\!\!&0.62&\!\!\!-\!\!\!&0.02&\!=\!& 2.17&Z=0.10
              \end{array}
       \right.
\end{eqnarray}
\begin{eqnarray}
  \label{Eq:DeltaTeffZ}
  \frac{\Delta \log T_{eff}}{\Delta \log Z} =
       \left\{ \begin{array}{cccccccr}
                 0.09&\!\!\!-\!\!\!&0.16&\!\!\!-\!\!\!&0.09&\!=\!&-0.16&\;\;\;\;\;Z=0.02\\
                 0.79&\!\!\!-\!\!\!&0.20&\!\!\!-\!\!\!&0.02&\!=\!& 0.57&Z=0.10
              \end{array}
       \right.
\end{eqnarray}
We note from Eq.~\ref{Eq:DeltaLZ} the small effect of the nuclear production
rate on the surface luminosity.

  These relations should not be considered very accurate
given the simplifications made in their derivation (mainly
because of the departure of the opacity profile from the Kramers law and the
resulting uncertainty in Eqs.~\ref{Eq:DeltaZ02} and \ref{Eq:DeltaZ10}). They
however enable to understand qualitatively the ZAMS location of a \mass{3} star
in the HR diagram as a function of the metallicity, and support
the following conclusions found in Fig.~\ref{Fig:HR}:

\vskip 1mm
\noindent - The surface luminosity at ZAMS is a decreasing function of $Z$
at $Z=0.02$ (and lower metallicities) and an increasing function of $Z$ at
$Z=0.1$.
The metallicity at which the inversion operates can only be determined from
complete stellar model calculations and is found to be $Z\simeq 0.05$.

\vskip 1mm
\noindent - The absolute shift $|\Delta \log L|$ resulting from an increase
$\Delta \log Z$
is much greater at $Z=0.1$ than at 0.02 (by a factor of about 14
according to Eq.~\ref{Eq:DeltaLZ}). This result is qualitatively well
verified by the ZAMS model calculations displayed in Fig.~\ref{Fig:HR}.

\vskip 1mm
\noindent - The effective temperature is a decreasing function of $Z$ for
$Z\simlr 0.05$, but it {\it increases} with metallicity for $Z\simgr 0.05$.

\end{document}